\documentclass[conference]{IEEEtran}
\IEEEoverridecommandlockouts
\usepackage{cite}
\usepackage{amsmath,amssymb,amsfonts}
\usepackage{algorithmicx}
\usepackage{algpseudocode}
\usepackage{algorithm}
\usepackage{graphicx}
\usepackage{amsmath}
\usepackage{amssymb}
\usepackage{array,booktabs, caption}
\usepackage{makecell}
\usepackage{longtable}
\usepackage{graphicx}
\usepackage{lipsum}
\usepackage{float}
\usepackage{multicol}
\usepackage{multirow}
\usepackage{textcomp}
\usepackage[table]{xcolor}
\usepackage{nicematrix}
\usepackage{blkarray}
\usepackage{xparse}
\usepackage{verbatim}
\usepackage{tabularx}
\usepackage{multirow}
\usepackage{bm}
\usepackage{stfloats}
\usepackage[caption=false,font=footnotesize]{subfig}
\usepackage{calc}
\usepackage{diagbox}
\usepackage{array}
\usepackage{multirow}
\usepackage{xcolor,graphicx,subfig,lipsum}
\usepackage{balance}

\newcolumntype{M}[1]{>{\centering\arraybackslash}m{#1}} 
\newcolumntype{P}[1]{>{\raggedright\arraybackslash}p{#1}} 

\def\BibTeX{{\rm B\kern-.05em{\sc i\kern-.025em b}\kern-.08em
    T\kern-.1667em\lower.7ex\hbox{E}\kern-.125emX}}
\begin{document}


\title{ZT-DAC: ML-Driven Dynamic Access Control Policy for Zero Trust in DER Networks}

\author{\IEEEauthorblockN{Arun B. Chandrasekaran, Damilola Popoola, Souradeep Bhattacharya, and Manimaran Govindarasu}
\IEEEauthorblockA{\textit{Department of Electrical and Computer Engineering} \\
\textit{Iowa State University, Ames, IA 50011}\\
Email: arun0810@iastate.edu, popson@iastate.edu, sbhatta@iastate.edu, gmani@iastate.edu
 }
}

\maketitle

\begin{abstract}
The increasing penetration of Distributed Energy Resources (DERs) in the smart grid has significantly expanded the cyber-attack surface. Traditionally, DER networks rely on static role-based access control (RBAC) mechanisms that assume persistent trust after authentication. This limitation enables adversaries to maintain unauthorized privileges and move laterally across DER systems, disrupting grid operations. Although attribute-based access control (ABAC) enables fine-grained and context-aware access decisions, it does not fully eliminate implicit trust. Zero Trust (ZT) principles can be incorporated to enhance ABAC policies. This paper proposes a dynamic access control framework for implementing ZT in DER networks (ZT-DAC). The proposed solution employs a two-stage verification process and integrates an Open Policy Agent (OPA)-based RBAC with a machine learning (ML)-based ABAC to enforce continuous and adaptive trust score assessment. To support realistic DER operations, a DER-specific access control model is developed and used to generate a domain-informed dataset of access requests incorporating both legitimate and adversarial behaviors. Experimental results on the dataset show that the OPA-based decision layer achieves \(91.92\%\) accuracy, with a False Positive Rate of \(14.45\%\), while the ML model achieves a Mean Absolute Error of \(2.074\) and an \(R^2\) of \(0.891\) in trust-score prediction. The ABAC layer achieves \(93.27\%\) accuracy, with a FPR of \(3.88\%\), validating its practical feasibility within realistic DER environments.

\end{abstract}


\vspace{-0.3cm}
\section{Introduction}
The modern power grid has evolved into a highly interconnected cyber-physical system (CPS), facilitated by the incorporation of Distributed Energy Resources (DERs), advanced metering infrastructure (AMI), edge intelligence, and cloud-based platforms \cite{10902093}. The integration of DERs, such as solar photovoltaics (PVs), wind farms, electric vehicles, and energy storage systems, offers several key advantages, such as enhancing the overall energy efficiency and resilience, providing ancillary services, and reducing the energy costs for consumers \cite{hahnder}. However, the growing involvement of utilities, vendors, consumers, and third-party service providers in DER ecosystems increases system complexity and expands the grid’s cyber-attack surface \cite{nerc2022dercyber}. DER networks have traditionally relied on perimeter-based security models, where entities inside the boundary are implicitly trusted \cite{11224995}. These architectures typically rely on conventional authentication and static identity- and role-based access controls (IBAC/RBAC), which grant persistent access after verification. However, their rule-based and non-adaptive policies provide limited awareness of user context and behavior. As a result, attackers who breach the perimeter can retain trusted access and move laterally across DER systems \cite{shtym2026eastwest}. Attribute-based access control (ABAC) can enhance static RBAC policies by incorporating contextual information such as user attributes, device state, location, and time of access. However, ABAC alone is insufficient to eliminate implicit trust or effectively prevent lateral movement within DER networks. To address this, Zero Trust (ZT) principles \cite{nist_sp_800_207} can be integrated with ABAC to enable continuous verification and adaptive access control, ensuring that every access request is dynamically evaluated based on identity, context, and trust conditions .



Considering the critical nature of DERs within the modern smart grid, we propose a ZT-enabled dynamic access control policy for DER networks (ZT-DAC). The proposed method employs a two-stage access request verification process implemented at the edge gateway. This objective is achieved through: \textbf{\textit{(i)}} a DER-specific access control model that captures stakeholder roles, permissions, and resource interactions; \textbf{\textit{(ii)}} a policy-based RBAC module implemented using Open Policy Agent (OPA) for role- and resource-level authorization; \textbf{\textit{(iii)}} a trust score formulation that quantifies request-level risk using behavioral and contextual features; and \textbf{\textit{(iv)}} an ML-based ABAC module that learns and refines trust estimates for adaptive access decisions. The framework is supported by a domain-informed dataset developed for this application and evaluated on a testbed environment with simulated DER devices, edge gateways, and DER users.


The remainder of the paper is divided as follows: Section II provides background on DER and ZT, highlighting DER architecture, cybersecurity challenges, the need for ZT in DER, and related work. Section III describes the proposed ZT-DAC methodology. Section IV discusses the experimental setup and dataset generation process. Section V discusses the experimental evaluation of the proposed method. Section VI concludes the paper with recommendations for future work.

\begin{table*}[t]
    \definecolor{lightgray}{RGB}{224,224,224}
	\renewcommand{\arraystretch}{1.3}
	\caption{Comparison of Proposed Work with Existing Literature in Access Control}
	\vspace{-0.2cm}
        \centering
	\label{tab:related_works_table}
	\centering
	\resizebox{\textwidth}{!}{
    
		\begin{tabular}{c c c c c c c}
		\hline\hline \\[-4mm]
        
		\multicolumn{1}{c}{\textbf{Method}} & \multicolumn{1}{c}{\textbf{OT/ICS}}  & \multicolumn{1}{c}{\textbf{DER-specific}} & \multicolumn{1}{c}{\textbf{Cybersecurity}} & \multicolumn{1}{c}{\textbf{Access Control}} & \multicolumn{1}{c}{\textbf{Machine Learning}} & \multicolumn{1}{c}{\textbf{Zero Trust}} \\ \hline

        Dynamic RBAC using RL \cite{9761965} & $\checkmark$ & $\checkmark$ & $\checkmark$ & $\checkmark$ & $\checkmark$ & $\times$ \\

        ZT Grid-Cloud AC \cite{10245681} & $\checkmark$ & $\times$ & $\checkmark$ & $\checkmark$ & $\times$ & $\checkmark$ \\

        BC-Resilient DERMS \cite{10795172} & $\checkmark$ & $\checkmark$ & $\checkmark$ & $\times$ & $\checkmark$ & $\times$ \\

        IPFS+BC Sharing \cite{10860565} & $\checkmark$ & $\checkmark$ & $\checkmark$ & $\checkmark$ & $\checkmark$ & $\times$ \\

        Adaboost AC for SCADA \cite{ZHOU2019548} & $\checkmark$ & $\checkmark$ & $\checkmark$ & $\checkmark$ & $\checkmark$ & $\times$ \\

        \rowcolor{lightgray}
        \textbf{ZT-DAC (Proposed)} & $\checkmark$ & $\checkmark$ & $\checkmark$ & $\checkmark$ & $\checkmark$ & $\checkmark$ \\

		\hline\hline
		\end{tabular}
	}
    \vspace{-0.4cm}
\end{table*}

\begin{figure}[t]
    \centering
    \includegraphics[width=7.5cm]{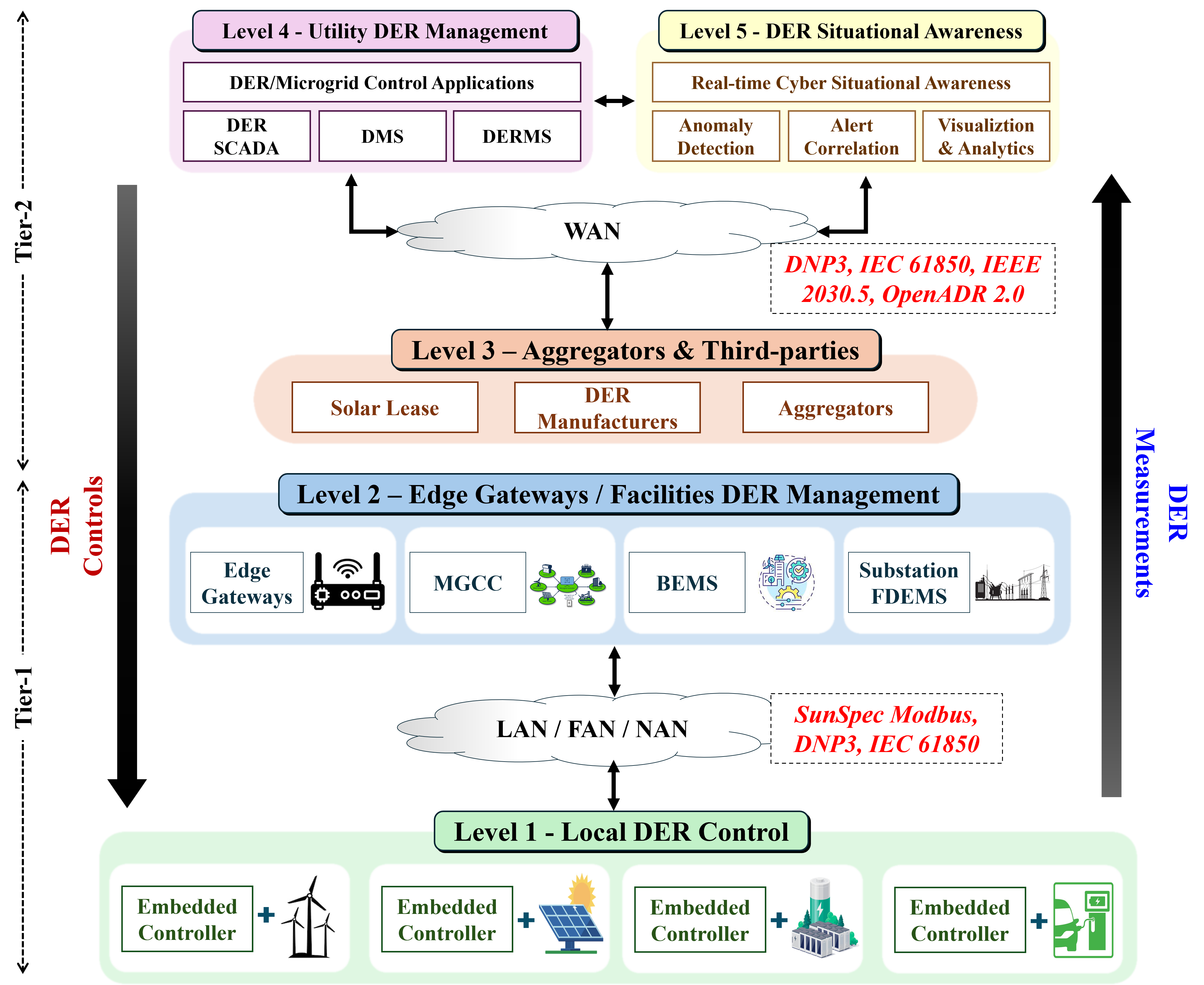}
    \caption{Five-level DER communication architecture}
    \label{fig:der_architecture}
    \vspace{-0.5cm}
\end{figure}

\section{Background on DER and Zero Trust}

\subsection{DER Architecture}

The DER communication architecture, as shown in Fig.~\ref{fig:der_architecture}, consists of multiple interconnected layers ranging from local DER controllers and edge gateways to aggregators, third-party service providers, utilities, and supervisory management platforms \cite{epri_2019}. At the lower layers, DER assets (solar PV, battery storage, EV chargers) communicate through local controllers and gateways using protocols such as SunSpec Modbus, DNP$3$, and IEC $61850$ \cite{10352402}. Higher layers integrate DER management systems (DERMS), aggregators, vendors, and utility operators. While this layered architecture enables large-scale monitoring, coordination, and control, it also introduces diverse stakeholders, communication interfaces, and trust relationships across the DER ecosystem \cite{osti_1765273}, highlighting the critical need for fine-grained access control mechanisms.


\vspace{-0.2cm}
\subsection{Cybersecurity Challenges and Need for Zero Trust in DER}


While DER connectivity improves operational efficiency, the integration of edge devices, cloud platforms, vendor-managed services, and customer-owned assets significantly expands the cyber attack surface \cite{hahnder}. Attackers who compromise exposed interfaces, weak credentials, or remote services can gain access to DER clients, gateways, smart inverters, sensors, and DERMS, potentially disrupting grid operations. Given the distributed and multi-stakeholder nature of DER networks, traditional perimeter-based security models that rely on implicit trust are no longer sufficient \cite{doe_DER_2022}. To address these risks, DER networks require a ZT approach that continuously verifies every request using contextual factors such as device posture, behavior, request frequency, time of access, and network risk indicators. This enables adaptive, fine-grained authorization and stronger resilience across modern DER ecosystems.

\vspace{-0.2cm}
\subsection{Overview of Zero Trust}

ZT offers a stronger security approach by enforcing the principle of \textit{``never trust, always verify"} \cite{nist_sp_800_207}. It is commonly implemented using two logical components: the Policy Decision Point (PDP) and the Policy Enforcement Point (PEP). The PDP is responsible for evaluating access requests against defined policies and contextual risk factors. It includes the Policy Engine (PE), which applies trust algorithms (TA) and decision logic, and the Policy Administrator (PA), which executes the resulting authorization decision. The PEP acts as the gatekeeper between requesting entities and protected resources, enforcing allow or deny actions issued by the PDP. 

\subsection{Related Work}

Existing research has focused on improving the security of DER-integrated grid through standards, vulnerability analysis, protocol security, anomaly detection methods, and situational intelligence \cite{epri_2019, doe_DER_2022, EPRI_NESCOR_DER_Security, DEBENEDETTI201859, 10744069, 10454168, CySIDER}. More recent studies have explored access-control mechanisms for DER environments \cite{9761965, 10795172, 10860565, ZHOU2019548}; however, many remain based on static authorization models or are not explicitly aligned with ZT principles. In parallel, research in ZT security has introduced ZT-specific architectures and implementations using network isolation, access control, and micro-segmentation strategies \cite{nist_sp_800_207, CISA_ZTMM_v2_2023, 9773102, 10013559, 10245681, 11224995}. While these efforts demonstrate the growing relevance of ZT, they are not tailored to the operational and multi-stakeholder DER networks. In particular, limited attention has been given to dynamic, context-aware access control that continuously evaluates trust across DER users, devices, and services. This gap motivates the need for a dynamic context-aware access control framework to enforce ZT principles in DER environments. Comparison table added in Table~\ref{tab:related_works_table} to address novelty and distinguish ZT-DAC from prior ZT access-control and DER security frameworks.

\begin{figure}[t]
    \centering
    \includegraphics[width=8cm]{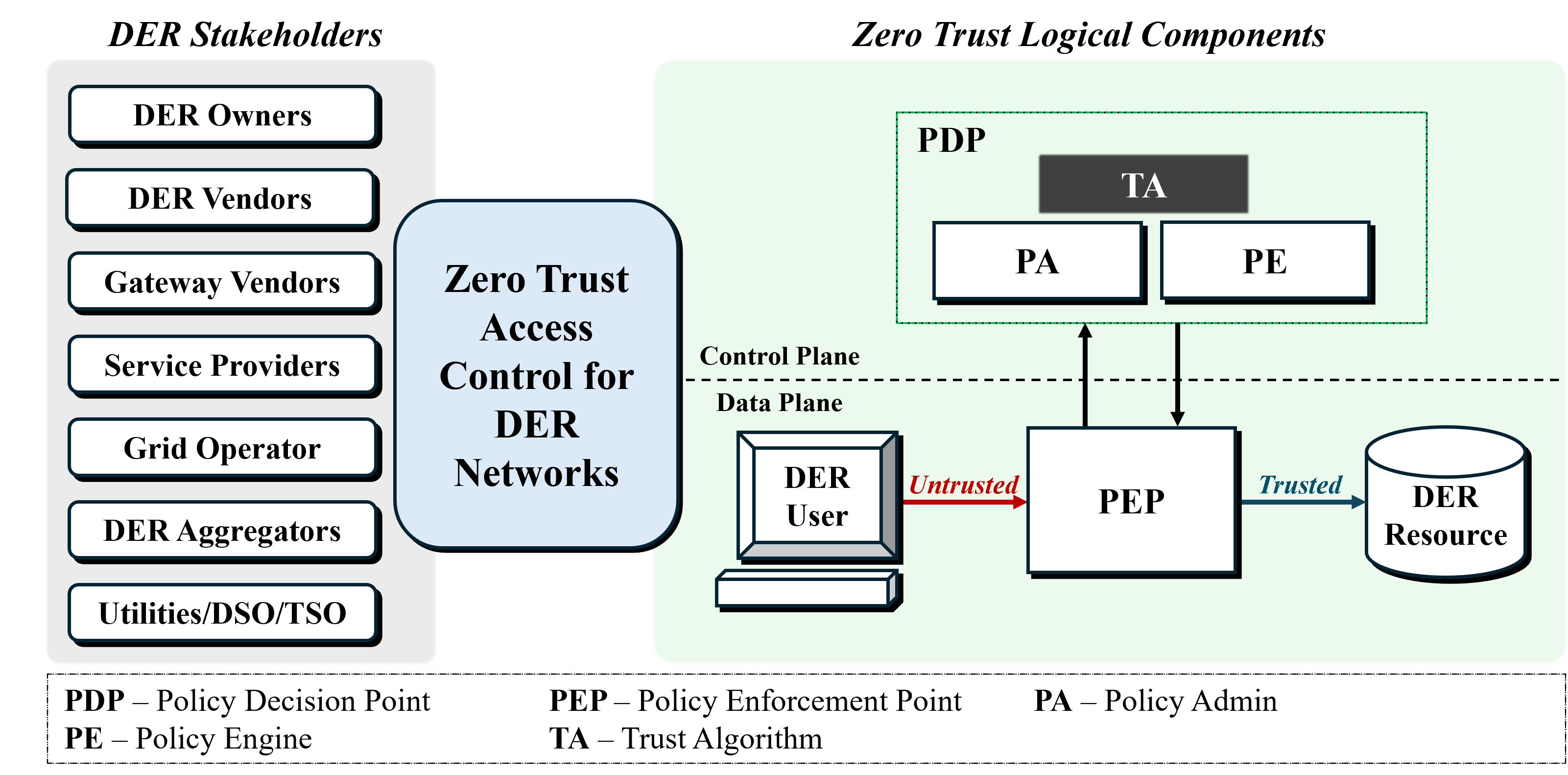}
    \caption{Conceptual overview of proposed ZT-DAC framework}
    \label{fig:dazed_framework}
    \vspace{-0.5cm}
\end{figure}


\section{Proposed ZT-DAC Framework}

An overview of the proposed ZT-DAC framework for smart grid DER networks is shown in Fig.~\ref{fig:dazed_framework}. The framework implements ZT by enforcing dynamic, user-level access control through the PDP-based Trust Algorithm and PEP. Incoming requests from DER users are treated as untrusted and are first filtered through an Access Control List (ACL) for coarse-grained identity validation. Valid requests are then evaluated at the PDP using a two-stage process that combines policy-based RBAC with ML-based ABAC and continuous trust-score assessment. The resulting decision is enforced at the PEP, which permits or denies access to DER resources. The major components of the framework are:


\subsection{DER Access Control Model}

DER environments involve multiple stakeholders with different roles, communication channels, and resource privileges. As a result, a uniform access policy is insufficient, since the same request may be legitimate for one entity but unsafe for another. To overcome this, a DER access control model, shown in Table~\ref{tab:access_control_table}, is developed using DER-specific guidelines from \cite{nerc2022dercyber, osti_1765273}. The proposed model maps DER roles to permitted actions, accessible resources, and cyber responsibilities to support context-aware authorization, providing the foundation for subsequent RBAC and trust-based evaluations.


\begin{table*}[t]
\renewcommand{\arraystretch}{1.15}
\caption{DER Access Control Model}
\label{tab:access_control_table}
\vspace{-0.2cm}
\centering
\footnotesize
\resizebox{\textwidth}{!}{
\begin{tabular}{
>{\raggedright\arraybackslash}p{3cm}
>{\raggedright\arraybackslash}p{3cm}
>{\raggedright\arraybackslash}p{4cm}
>{\raggedright\arraybackslash}p{2.5cm}
>{\raggedright\arraybackslash}p{4cm}
}
\hline\hline
\textbf{Role} & \textbf{Ownership} & \textbf{Access Channel} & \textbf{Grid Actions} & \textbf{Cyber Responsibilities} \\
\hline

DER Owner (Residential) & DER equipment & Local interfaces (Wi-Fi, Zigbee) & None & Delegates control to vendor/aggregator \\

DER Owner (Commercial) & DER equipment & Vendor portals or cellular (4G/5G) & None & DER management \\

DER Vendor (OEM) & None (post-deployment) & Proprietary communication channels & None & Firmware patching (on request) \\

Gateway Vendor & None (post-deployment) & None & None & Gateway patching and updates \\

DER Service Provider & None & APIs and DER control servers (standardized/proprietary) & None & Firmware updates and access control management \\

DER/Gateway Installer & None (post-deployment) & Local interfaces (Wi-Fi, Zigbee) & None & Device configuration and security setup \\

Grid Operator & Grid infrastructure, DERMS & IEEE 2030.5, DNP3, SunSpec, vendor portals & Full operational control & Compliance with NERC CIP \\

DER Aggregator & Control servers, DER APIs & Standardized protocols or vendor portals & Utility-delegated actions & Infrastructure and API protection \\

DSO & Distributed system control & Indirect via utilities/aggregators & Distribution grid management & Compliance with NERC CIP \\

\hline\hline
\end{tabular}
}
\vspace{-0.3cm}
\end{table*}

\begin{figure}[t]
    \centering
    \includegraphics[width=7.5cm]{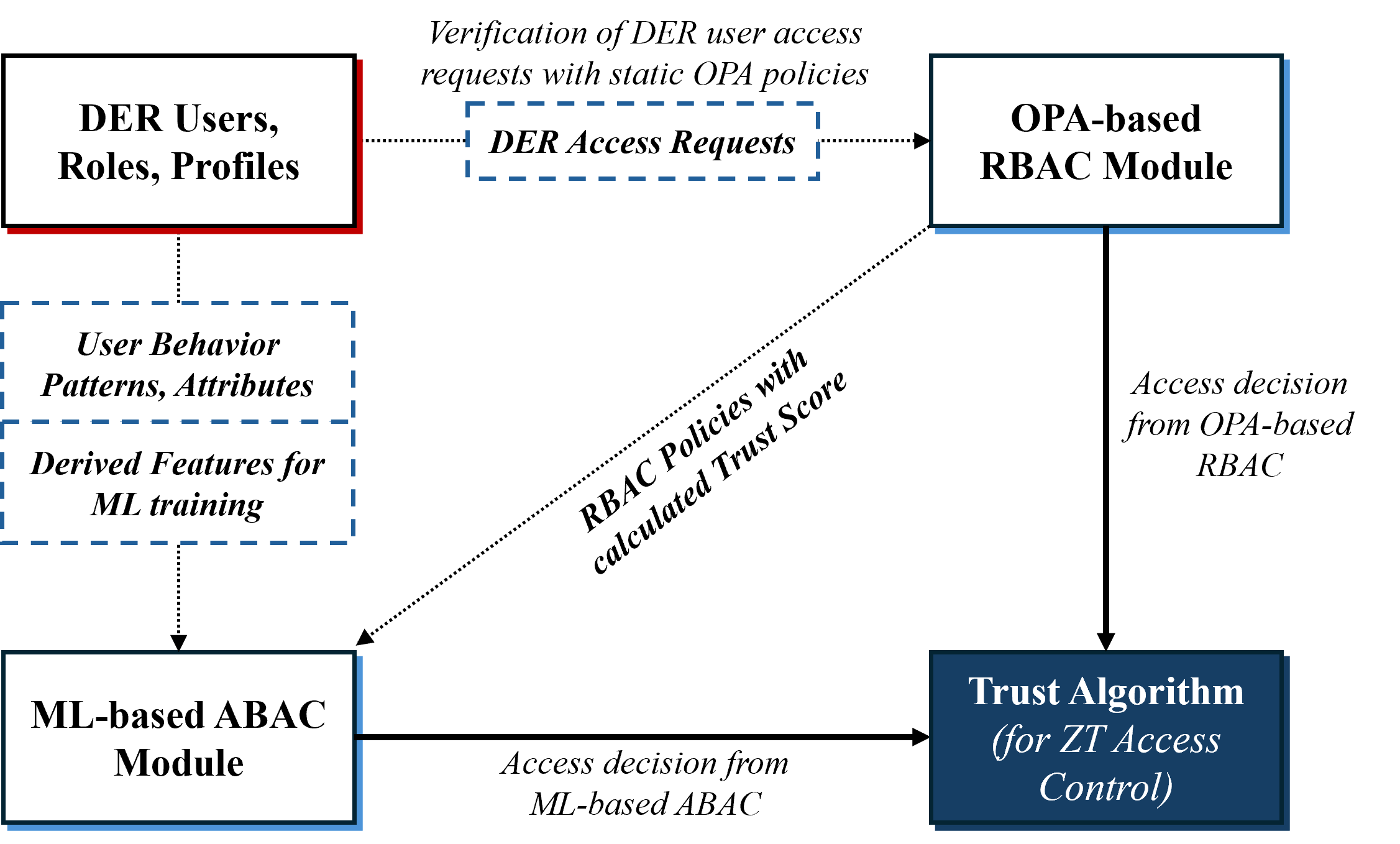}
    \caption{ZT-DAC PDP framework}
    \label{fig:pdp_framework}
    \vspace{-0.5cm}
\end{figure}

\begin{algorithm}[ht]
\caption{OPA-RBAC Top-Level Policy Evaluation}
\footnotesize
\begin{algorithmic}[1]

\Require Request $r=\langle role,action,res,dev,user\rangle$
\Require Policy data $\mathcal{P}$, time $t$, threshold $\tau$
\Ensure \textbf{ALLOW} or \textbf{DENY}

\If{
    \textbf{not }ValidRoleAct(role,action) $\lor$
    \textbf{not }ValidRes(role,res) $\lor$
    \textbf{not }DevHealthy(dev) $\lor$
    \textbf{not }CertOK(dev.cert,t) $\lor$
    ThreatFlag(user.id,$\mathcal{P}$) $\lor$
    \textbf{not }TrustOK(user.ts,$\tau$)
}
    \State \textbf{return DENY}
\EndIf

\State \textbf{return ALLOW}

\end{algorithmic}
\vspace{0.1cm}
\end{algorithm}

\subsection{PDP for Dynamic Access Control}
An overview of the PDP is shown in Fig.~\ref{fig:pdp_framework} and it includes three components as discussed below:

\subsubsection{OPA-based RBAC}

The first authorization stage uses OPA, a policy-as-code engine that evaluates structured JSON requests. It separates decision logic from enforcement, enabling deterministic, policy-based validation of each DER access request. As described in Algorithm~1, each request is encoded with identity, role, action, resource, and device attributes, following Table~\ref{tab:access_control_table}. Rego policies enforce RBAC and ACL constraints, including \textit{role--action--resource} consistency, certificate validity, and basic trust checks. This stage computes a rule-based trust score for OPA ($\mathbf{TS_1}$), given by Eqn.~\ref{eqn:ts1}, using four normalized behavioral sensitivity indicators: \textit{(i)} request-rate $(r_{\mathrm{req}})$ (frequency of access requests), \textit{(ii)} recent-denial $(r_{\mathrm{rev}})$ (recent revoked requests), \textit{(iii)} access-time $(r_h)$ (time-of-day access behavior), and \textit{(iv)} IP reputation $(r_{\mathrm{ip}})$ (network trustworthiness). Here, $S\rightarrow$ Weighted Sensitivity and $w_i\rightarrow$ weights (normalized to sum to one and can be adjusted based on system requirements). We have considered the following: $w_{\mathrm{req}}=0.30$, $w_{\mathrm{rev}}=0.30$, $w_h=0.20$, $w_{\mathrm{ip}}=0.20$.
\begin{equation}
\mathbf{TS_1} = 100(1-S), \quad TS_1 \in [0,100]
\label{eqn:ts1}
\vspace{-0.3cm}
\end{equation}

\vspace{-0.2cm}
\begin{equation}
S =
w_{\mathrm{req}}r_{\mathrm{req}} +
w_{\mathrm{rev}}r_{\mathrm{rev}} +
w_h r_h +
w_{\mathrm{ip}}r_{\mathrm{ip}}
\end{equation}

 The final decision at this stage is \textbf{ALLOW} only if all policy conditions are satisfied and $TS_1 \geq 85$; otherwise, \textbf{DENY}.





\begin{figure}[t]
    \centering
    \includegraphics[width=8.5cm]{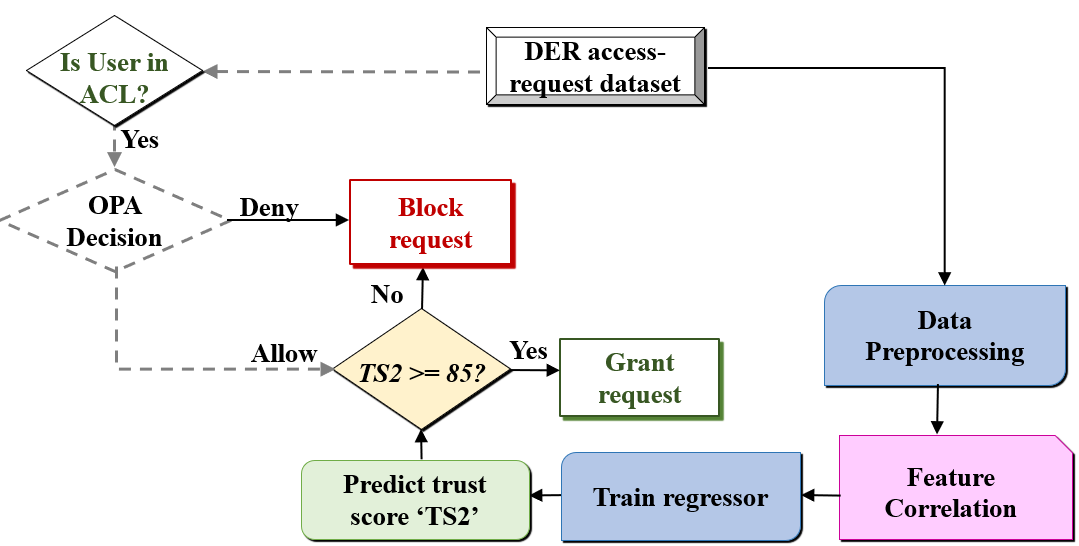}
    \caption{OPA-ML Integrated Workflow}
    \label{fig:ML_Workflow}
    \vspace{-0.6cm}
\end{figure}

\subsubsection{ML-based ABAC}
\label{sec:ml_abac}

The ML-based ABAC module complements the OPA-based RBAC layer by providing adaptive trust evaluation through a regression-based trust prediction, as illustrated in Fig.~\ref{fig:ML_Workflow}. The ML module predicts a continuous trust score ($\mathbf{TS_2}$) for each DER access request using contextual, behavioral, and device-level features. A custom-generated DER access-request dataset is used for training and testing. First, data preprocessing is performed to clean inconsistent entries, handle missing values, normalize numerical features, and convert the raw JSONL data into a structured format suitable for ML training. Next, feature correlation analysis using \textit{Pearson correlation} identifies the most relevant features influencing trust evaluation while reducing redundancy. The processed dataset is then used to train three regression models: Histogram-based Gradient Boosting (HGB), Random Forest (RF), and Light Gradient Boosting Machine (LGBM). The regressor learns a mapping function to input feature vector ($\mathbf{x}_t$) for request $t$:
\vspace{-0.15cm}
\begin{equation}
\hat{TS}_2(t) = f_\theta(\mathbf{x}_t); \quad \hat{TS}_2(t) \in [0,100]
\end{equation}

 The regression target is a reference trust score produced by a deterministic trust-evaluation component before policy evaluation. The ML model is not provided with the scoring equation, component weights, or normalized intermediate risk values. It receives only the observable request attributes and behavioral history derived from the request stream, and accordingly, the regression evaluation measures the model’s ability to approximate and generalize the reference trust-scoring mechanism.

\subsubsection{ZT-enabled Trust Algorithm}

The proposed ZT-DAC combines OPA-RBAC with ML-ABAC to employ a two-stage dynamic access control policy. Initial access is granted only if both trust scores satisfy the required threshold:
\vspace{-0.1cm}
\begin{equation}
\text{Access}(r_t)=
\begin{cases}
\textsc{allow}, & TS_1 \geq 85 \land TS_2 \geq 85 \\
\textsc{deny}, & \text{otherwise}
\end{cases}
\end{equation}

After access is granted, the ML-ABAC module continuously re-evaluates $\mathbf{TS_2}$ for every new request within an existing session. The updated trust context is derived from behavioral indicators ($r_{\mathrm{req}}$ and $r_{\mathrm{rev}}$). If the predicted trust score falls below the threshold ($TS_2 < 85$), access is immediately revoked. By continuously validating user trustworthiness, the framework enforces ZT in DER environments. The advantages of the proposed ZT-DAC over conventional access control are highlighted in Table~\ref{tab: ac_comparison}.


\begin{table}[t]
\caption{Comparison of Access Control Techniques}
\vspace{-0.35cm}
\label{tab: ac_comparison}
\begin{center}
\begin{tabular}{ >{\centering\arraybackslash}m{1.5in} >{\centering\arraybackslash}m{0.4in} >{\centering\arraybackslash}m{0.4in} >{\centering\arraybackslash}m{0.48in}}
\hline\hline

\textbf{Properties} & \textbf{RBAC} & \textbf{ABAC} & \textbf{ZT-DAC}\\ \hline
    
Role-Based Authorization & $\checkmark$ & $\checkmark$ & $\checkmark$ \\ 

Context-Aware Authorization & $\times$ & $\checkmark$ & $\checkmark$ \\ 

Behavior-Aware Trust Evaluation & $\times$ & $\checkmark$ & $\checkmark$ \\ 

Continuous Trust Re-evaluation & $\times$ & $\times$ & $\checkmark$ \\ 

Adaptive Access Control & $\times$ & $\times$ & $\checkmark$ \\ 

Lateral Movement Prevention & $\times$ & $\times$ & $\checkmark$ \\ 

Zero Trust Enforcement & $\times$ & $\times$ & $\checkmark$ \\ \hline\hline

\end{tabular}
\end{center}
\vspace{-0.3cm}
\end{table}

\vspace{-0.1cm}
\subsection{NGFW-based PEP}

The PEP operates at the network boundary to enforce the authorization decisions generated by the PDP. Requests satisfying the trust and policy conditions are permitted, while unauthorized requests are blocked before accessing protected DER services. In practical deployments, the PEP can be implemented using a Next-Generation Firewall (NGFW) due to its support for traffic filtering, identity-aware inspection, and centralized policy enforcement. However, NGFW-based PEP implementation is beyond the scope of this work and will be incorporated in future enhancements.

\begin{figure}[t]
    \centering
    \includegraphics[width=8.5cm]{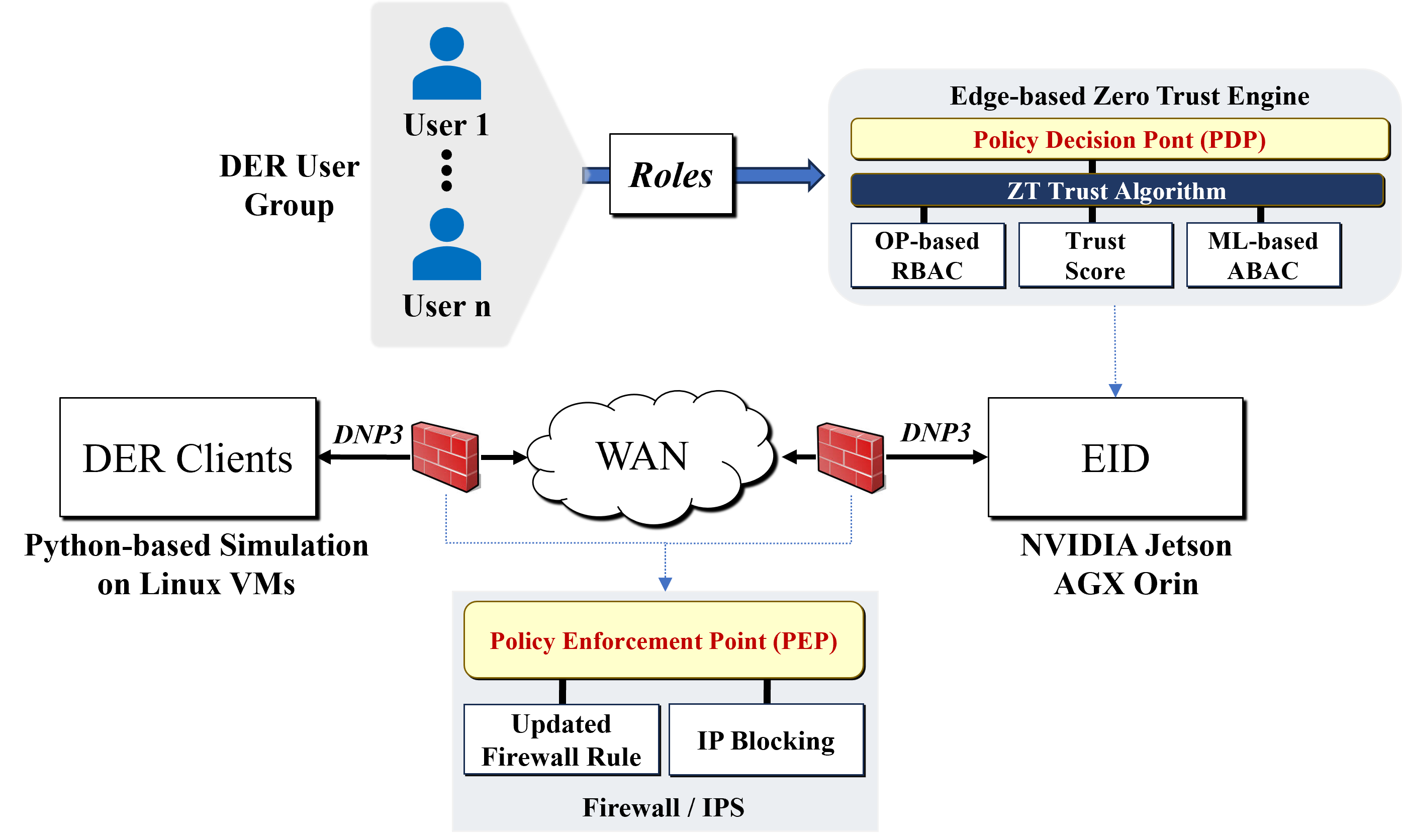}
    \caption{Experimental setup for implementation of ZT-DAC}
    \label{fig:experimental_testbed}
    \vspace{-0.2cm}
\end{figure}

\section{Experimental Setup \& Dataset Generation}

\subsection{Cyber-Physical Testbed Environment}

The experimental setup, shown in Fig.~\ref{fig:experimental_testbed}, is implemented using the PowerCyber Testbed at Iowa State University (ISU) \cite{7781277}. The testbed emulates a DER communication environment where user requests are evaluated by the proposed ZT-DAC framework before accessing DER resources. VMware vCenter platform is used to deploy Linux-based virtual machines representing DER clients and user entities, while DER DNP$3$ communication is emulated using the \texttt{dnp3-python} library. An NVIDIA Jetson AGX Orin functions as the edge gateway hosting the PDP, through which all DER access requests are routed.  Although the proposed architecture envisions an NGFW-based PEP, implementing a production-grade NGFW is beyond the scope of this work. To realize an end-to-end proof-of-concept and complete the policy enforcement loop, a pfSense firewall is configured as the PEP in the experimental testbed to enforce the access-control decisions. The use of pfSense is solely for prototype validation and is not presented as an NGFW implementation.

\begin{table*}[t]
\centering
\caption{Representative legitimate and illegitimate DER access-request scenarios}
\label{tab:request_scenarios}
\renewcommand{\arraystretch}{1.15}
\setlength{\tabcolsep}{4pt}

\begin{tabularx}{\textwidth}{
|>{\centering\arraybackslash\setlength{\parindent}{0pt}}X
|>{\centering\arraybackslash\setlength{\parindent}{0pt}}m{0.34\textwidth}
|>{\centering\arraybackslash\setlength{\parindent}{0pt}}X|
}
\hline
\textbf{Scenario} &
\textbf{Attributes} &
\textbf{Legitimate/Illegitimate}
\\
\hline

Grid operator requests \texttt{monitor\_grid\_cyber\_events} for an authorized DERMS resource.
&
&
Legitimate
\\
\cline{1-1}\cline{3-3}

DER aggregator requests \texttt{dispatch\_der\_resources} for an assigned DER fleet.
&
\multirow{3}{=}{
Healthy device, compliant firmware, valid certificate,
normal request rate, expected operating hour, and low
IP-reputation risk.}
&
Legitimate
\\
\cline{1-1}\cline{3-3}

&
&
\\

Residential DER owner requests \texttt{issue\_voltage\_setpoints}, an operation reserved for the grid operator.
&
&
Illegitimate

\\
\hline

DER service provider requests \texttt{rotate\_credentials} on an authorized gateway using an expired certificate.
&
Healthy device, compliant firmware, expired certificate,
authorized role--action--resource mapping, normal request rate,
expected operating hour, and low IP-reputation risk.
&
Illegitimate
\\
\hline
\end{tabularx}
\vspace{-0.4cm}
\end{table*}

\begin{table}[t]
\centering
\caption{DER Access-Control Dataset Features}
\label{tab:dataset_features}
\vspace{-0.2cm}
\renewcommand{\arraystretch}{1.1}
\footnotesize
\begin{tabular}{>{\centering\arraybackslash}p{3cm}>{\centering\arraybackslash}p{5cm}}
\hline\hline
\textbf{Feature} & \textbf{Description} \\
\hline
client\_id & Unique identifier of requesting entity. \\

allow & Ground-truth access decision. \\

role & Operational role of requester. \\

action & Requested operation. \\

resource & Target DER asset or service. \\

health & Device health state. \\

fw & Firmware version indicator. \\

hour & Request hour (0--23). \\

ip\_rep & Normalized IP reputation score. \\

is\_healthy & Encoded device health status. \\

fw\_ok & Firmware compliance indicator. \\

trust\_score & Baseline trust score of the request. \\
\hline\hline
\end{tabular}
\vspace{-0.2cm}
\end{table}

\begin{table}[t]
\centering
\caption{Evaluation Metrics and Descriptions}
\vspace{-0.2cm}
\label{tab:evaluation_metrics}
\footnotesize
\renewcommand{\arraystretch}{1.2}
\begin{tabular}{|>{\centering\arraybackslash}m{0.28\linewidth}|>{\centering\arraybackslash}m{0.64\linewidth}|}
\hline
\textbf{Metrics} & \textbf{Description} \\
\hline
Mean Absolute Error (MAE) & Avg. difference between predicted and reference trust scores. \\
\hline
Root Mean Squared Error (RMSE) & Prediction error with higher penalty for large deviations. \\
\hline
Coefficient of Determination ($R^2$) & Variance in trust score explained by the model. \\
\hline
Accuracy & Percentage of correct allow/deny decisions. \\
\hline
False Positive Rate (FPR) & Legitimate requests incorrectly denied. \\
\hline
False Negative Rate (FNR) & Suspicious requests incorrectly allowed. \\
\hline
\end{tabular}
\vspace{-0.5cm}
\end{table}

\subsection{DER Access-Control Dataset Generation}

\begin{figure}[t]
    \centering
    \includegraphics[width=9.0cm]{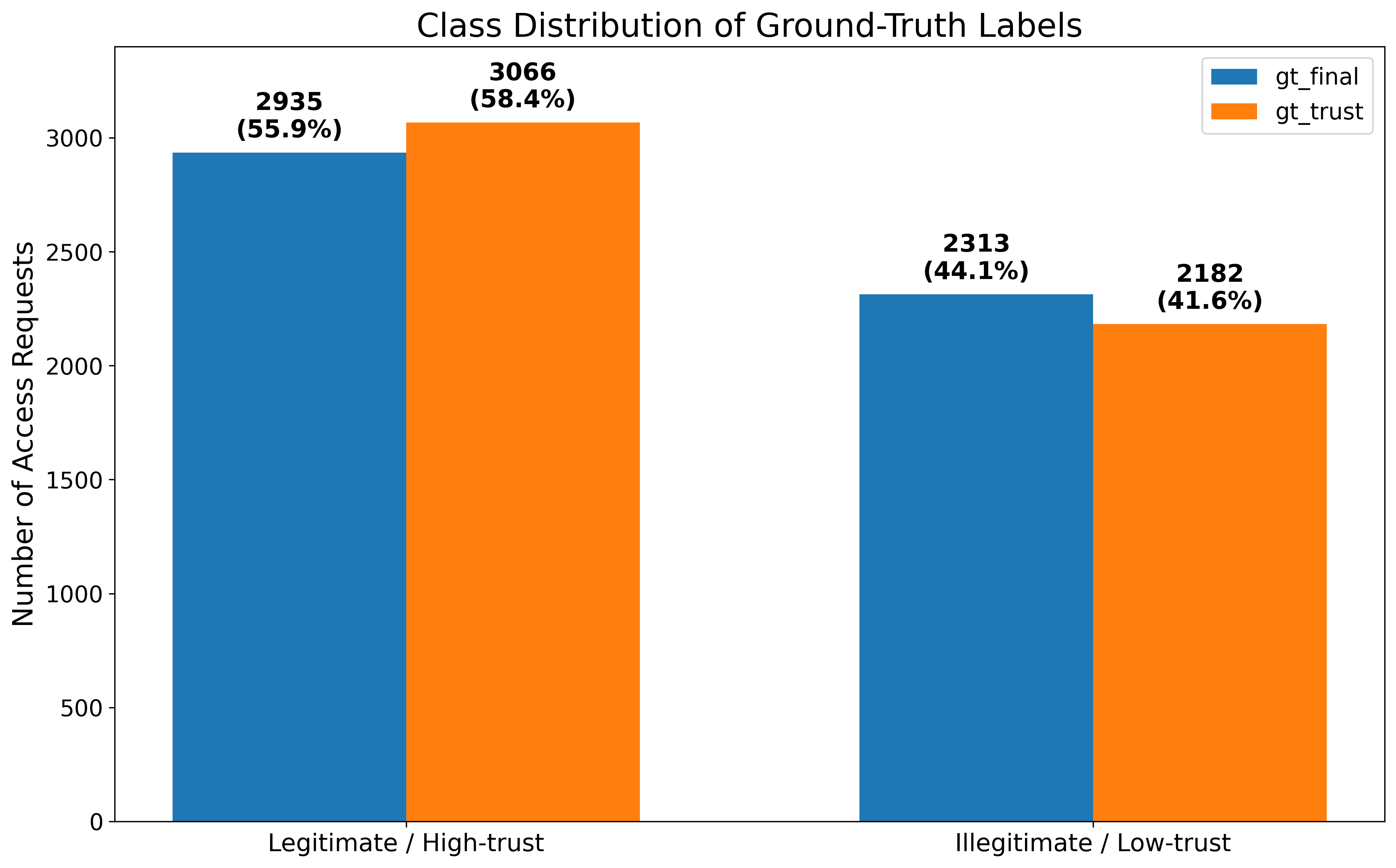}
    \caption{Class Distribution based on each ground truth label}
    \label{fig:class_bal}
    \vspace{-0.2cm}
\end{figure}

To evaluate the proposed ZT-DAC framework, a synthetic DER access-request dataset is generated based on domain-specific roles. An RBAC table is constructed from Table~\ref{tab:access_control_table}, defining valid \textit{role--action--resource} mappings \cite{nerc2022dercyber, osti_1765273}. Each record represents a request, where legitimate samples follow valid mappings, and illegitimate samples introduce policy violations. Roles exhibit distinct request patterns, while actions reflect expected operational behavior. Each request includes contextual attributes such as device health, firmware status, request time, and IP reputation. The dataset captures both normal and anomalous access scenarios and is exported in JSON format (sample scenarios can be viewed in Table~\ref{tab:request_scenarios}, along with the class balance in Fig.~\ref{fig:class_bal}). The features in the dataset are shown in Table~\ref{tab:dataset_features}. Dataset will be made public in future.

\section{Experimental Evaluation}
\label{sec:exp_eval}

\subsection{Case Study Scenarios}

The proposed ZT-DAC was evaluated through two case studies corresponding to the major stages of the PDP pipeline: \textit{(1) ML-based trust prediction:} assesses the ability of the ML-based ABAC module to predict contextual trust scores ($TS_2$) using behavioral, contextual, and device-related features, and \textit{(2) End-to-end ZT-DAC PDP evaluation:} evaluates the complete ZT-DAC PDP by integrating OPA-RBAC and ML-ABAC to generate final allow/deny access decisions under realistic DER request conditions.

\subsection{Results and Discussion}
The proposed ZT-DAC is evaluated using the DER access-request dataset and performance metrics provided in Table~\ref{tab:evaluation_metrics}. 



\subsubsection{ML-based trust prediction}

\begin{figure}[t]
\centering
\includegraphics[width=8cm]{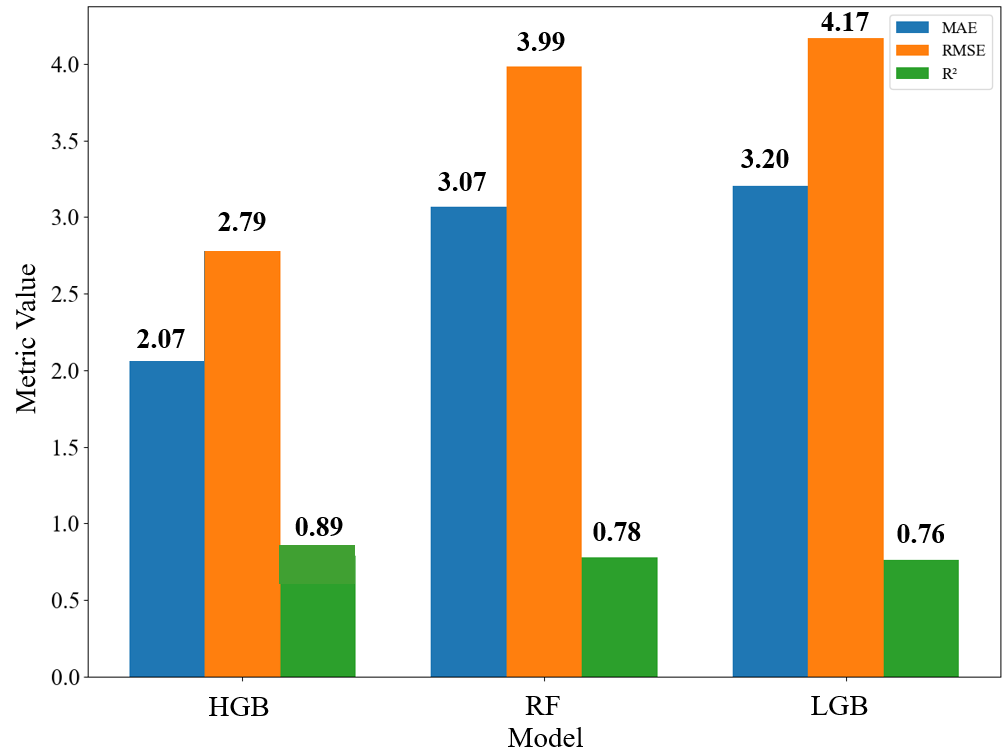}
\caption{Offline performance comparison of ML-ABAC regression models on DER access-request dataset}
\label{fig:trust_model_performance}
\vspace{-0.2cm}
\end{figure}

\begin{figure}[t]
\centering
\includegraphics[width=8cm]{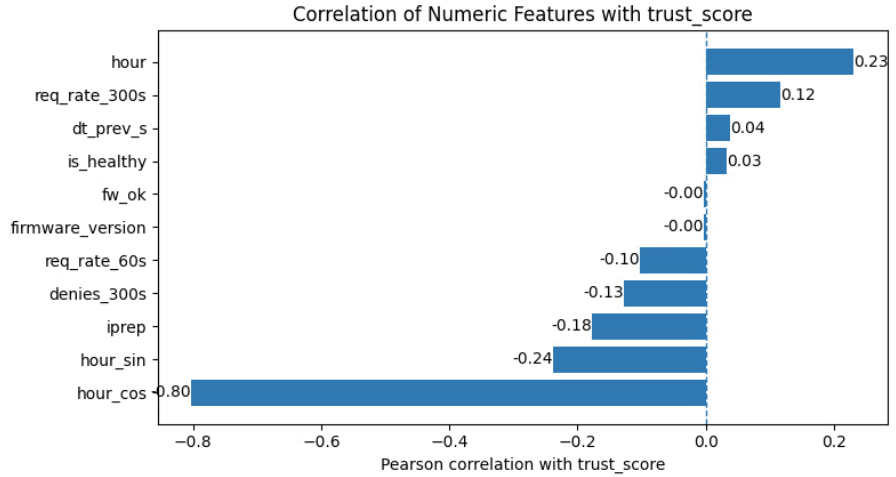}
\caption{Feature correlation for selected HGB model}
\label{fig:feature_correlation}
\vspace{-0.45cm}
\end{figure}

This case study evaluates the offline performance of the ML models. A total of $5248$ records were used, with a $70-30$ split for training and validation. Three tree-based models (HGB, RF, and LGB) were evaluated and the results are shown in Fig.~\ref{fig:trust_model_performance}. HGB achieved the best overall performance with \(\mathrm{MAE}=2.07\), \(\mathrm{RMSE}=2.79\), and \(R^2=0.89\), outperforming RF \((\mathrm{MAE}=3.07,\ \mathrm{RMSE}=3.99,\ R^2=0.78)\) and LGB \((\mathrm{MAE}=3.20,\ \mathrm{RMSE}=4.17,\ R^2=0.76)\). The feature correlation analysis for HGB (especially the negative correlation) as shown in Fig.~\ref{fig:feature_correlation}, indicates that increasing values of a feature are associated with decreasing trust under the reference trust formulation. For example, higher IP-reputation risk, odd hours, recent denial history, and elevated request rates increase modeled behavioral risk and therefore have the strongest influence on the predicted $\mathbf{TS_2}$. These results indicate that the ML-based ABAC module can reliably learn contextual trust behavior from DER access-request patterns and accurately estimate trust scores for subsequent access-control evaluation. Based on the results, HBG was selected for end-to-end implementation.



\begin{figure}[t]
\centering
\includegraphics[width=7.5cm]{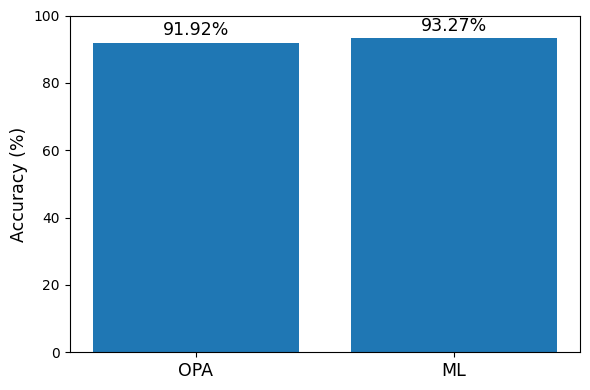}
\caption{Decision accuracy of the OPA-based RBAC and the ML-based ABAC with respect to their corresponding ground-truth labels.}
\label{fig:acc_plot}
\vspace{-0.2cm}
\end{figure}

\begin{figure}[t]
\centering
\includegraphics[width=8cm]{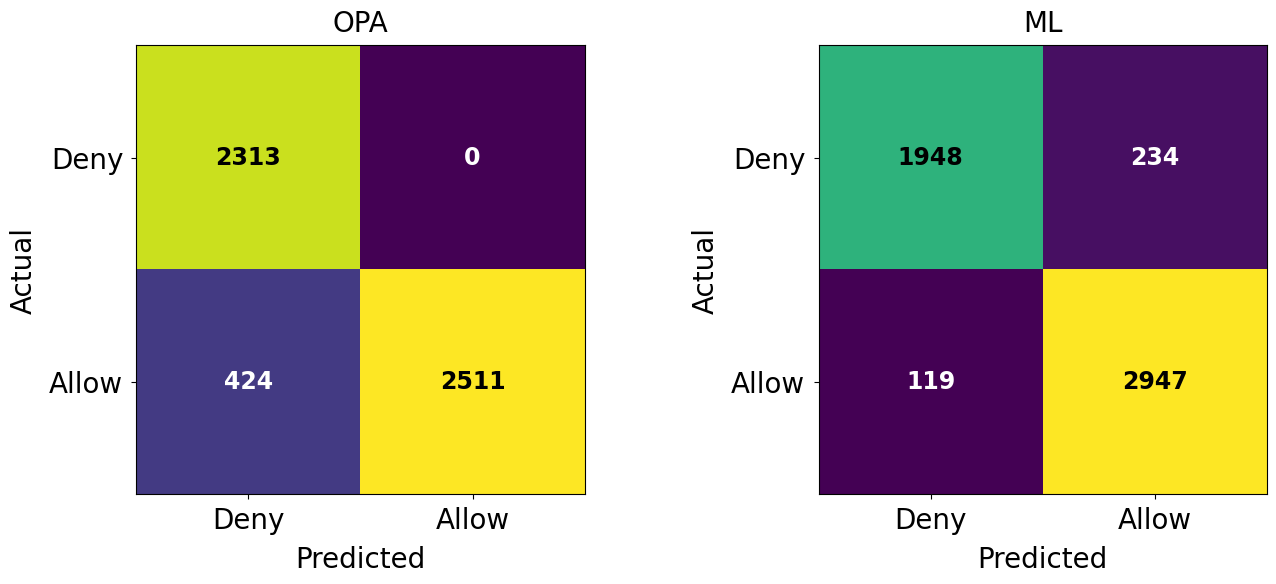}
\caption{Decision-level confusion matrices for OPA-based policy evaluation and ML-based trust evaluation.}
\label{fig:conf_mat}
\vspace{-0.5cm}
\end{figure}

\begin{table}[t]
\centering
\caption{Decision Latency Results}
\label{tab:latency_results}
\begin{tabular}{lrrrr}
\hline
\textbf{Decision path} &
\textbf{Mean} &
\textbf{Median} &
\textbf{Min} &
\textbf{Max} \\
\hline
OPA path & 5.576 ms & 4.550 ms & 3.150 ms & 20.530 ms
\\
ML-only  & 3.163 ms & 3.060 ms & 2.363 ms & 10.506 ms 
\\
\hline
\end{tabular}
\vspace{-0.5cm}
\end{table}

\subsubsection{End-to-end decision evaluation}

This case study evaluates the complete ZT-DAC PDP workflow by integrating the OPA-based RBAC module with the HGB-based ML trust predictor in the experimental testbed. The testing set access requests data were streamed into the PDP engine, deployed in the edge device, to emulate realistic DER communication traffic and dynamic access-control conditions. For each request, the OPA layer first performed policy validation and rule-based trust evaluation ($TS_1$), after which the ML-based ABAC module predicted the contextual trust score ($TS_2$). Final access decisions were then generated using the proposed two-stage trust evaluation mechanism.

The latency results in Table~\ref{tab:latency_results} characterize the computational overhead of the two decision components separately rather than the complete end-to-end PDP--PEP enforcement delay. The recorded OPA decision path achieved a mean latency of 5.576 ms per request, with a median of 4.550 ms. The HGB-based ML trust predictor achieved a lower mean latency of 3.163 ms, with a median of 3.060 ms.

The OPA-based RBAC layer achieved an accuracy of \(91.92\%\), with
\(TP=2313\), \(TN=2511\), \(FP=424\), and \(FN=0\), corresponding to an
FPR of \(14.44\%\) and an FNR of \(0\%\), as shown in
Fig.~\ref{fig:acc_plot} and Fig.~\ref{fig:conf_mat}. The absence of FN and its comparatively high FPR reflects a security-availability tradeoff, wherein, stricter enforcement reduces the likelihood of unauthorized access but may delay or block legitimate DER operations. This can be mitigated through periodic policy calibration using representative operational data.

The ML-based ABAC module achieved an accuracy of \(93.27\%\), with
\(TP=1948\), \(TN=2947\), \(FP=119\), and \(FN=234\), corresponding to an
FPR of \(3.88\%\) and an FNR of \(10.73\%\). The lower FPR indicates greater flexibility in evaluating contextual and behavioral attributes, whereas the higher FNR reflects cases in which illegitimate requests received predicted
trust scores above the authorization threshold. Since the decision process is sequential, an ML-based approval cannot override an earlier OPA denial, as, the final protection against ML false negatives depends on whether those requests are also rejected by the OPA layer.
These results demonstrate the complementary roles of deterministic policy enforcement and context-aware trust estimation in ZT-DAC.

\section{Conclusion \&\ Future Work}
This paper presented ZT-DAC, a policy-based and ML-driven dynamic access-control framework for implementing ZT in DER networks. By combining OPA-based RBAC with ML-based ABAC, the framework evaluates both rule-based trust score ($\mathbf{TS_1}$) and ML-predicted contextual trust score ($\mathbf{TS_2}$), arriving at a unified decision to grant or deny DER access requests. Experimental results on a custom DER-specific access-control dataset show that the OPA-RBAC layer achieved an accuracy of $91.92\%$ with an FPR of $14.44\%$, while the ML-ABAC module achieved a higher accuracy of $93.27\%$ and reduced the FPR to $3.88\%$. These results demonstrate the effectiveness of the proposed ZT-DAC in enabling continuous and context-aware ZT access control for DER environments. Future work includes: \textbf{\textit{(i)}} implementing NGFW-based PEP for policy enforcement, \textbf{\textit{(ii)}} real-time testing and validation using a hardware-in-the-loop DER testbed, and \textbf{\textit{(iii)}} enhancement of adaptive policy and trust-decision optimization using reinforcement learning.


\section*{Acknowledgement}
This work is funded in part by the US DOE Cybersecurity, Energy Security, and Emergency Response (CESER) Award Number DE-CR0000049.

\bibliographystyle{IEEEtran}
\bibliography{RefDatabase}

\end{document}